\documentclass{edp-jp4}
\usepackage{graphicx}
\def\lbcoate{La$_{1.875}$Ba$_{0.125}$CuO$_4$}

\def\lsno{La$_{2-x}$Sr$_x$NiO$_4$}

\def\lsco{La$_{2-x}$Sr$_x$CuO$_4$}

\def\ybco{YBa$_2$Cu$_3$O$_{6+x}$}

\begin{document}

\title{Charge stripes in cuprate superconductors: The middle way} 
\author{J. M. Tranquada}
\address{Brookhaven National Laboratory, Upton, NY 11973-5000, USA}
\maketitle
\begin{abstract} Charge and spin stripe order is a type of electronic
crystal observed in certain layered cuprates associated with
high-temperature superconductivity.  Quantum-disordered stripes could be
relevant for understanding the superconductivity.  Here I discuss recent
experimental characterizations of the stripe-ordered state in \lbcoate,
and compare them with properties of superconducting compositions.
\end{abstract}
%

\section{Introduction}

For two-dimensional (2D) electronic systems, there are two limiting
ground states.  When kinetic energy dominates, one obtains a Fermi
liquid, with delocalized quasiparticles.  On the other hand, when Coulomb
repulsion dominates, one ends up with electrons localized in a Wigner
crystal \cite{baer05}.  The parent compounds of cuprate superconductors
are Mott-Hubbard charge-transfer insulators, where neither of the limiting
ground states is applicable. The doped antiferromagnetic insulator
becomes metallic and superconducting, with a normal state that is not
well understood.  There have been attempts to describe various aspects of
doped cuprates starting with either a Fermi-liquid picture \cite{chub03}
or a Wigner crystal description \cite{fran04,rast04,komi05}, though there
are features of each that are not entirely satisfactory.

As discussed at this workshop by Boris Spivak, there is also a ``middle
way''.  A proper treatment of the transition from a Fermi liquid state to
a Wigner crystal indicates that there should be intermediate
``micro-emulsion'' phases \cite{jame05}.  In the case of hole-doped
antiferromagnets such as cuprates, these are predicted to be stripe
phases \cite{kive03,zaan01,mach89,sach91}.  Indeed, ordered charge and
spin stripes have been observed in certain cuprates (especially near a
hole concentration of $\frac18$ per Cu site) \cite{ichi00,fuji04} and in
layered nickelates over a broad range of doping \cite{yosh00}.  Static
stripe order competes with superconductivity \cite{ichi00}, but
quantum-disordered stripes could be compatible, and perhaps responsible
for, high-temperature superconductivity \cite{emer97,arri04}.

In this paper, I will discuss some recent results for \lbcoate, a compound
that exhibits static stripe order at temperatures below $\approx 50$~K
\cite{fuji04}.  The magnetic excitation spectrum in the ordered state
\cite{tran04} looks very much like that in superconducting \ybco\
\cite{hayd04}.  New measurements on \lbcoate\ in the paramagnetic state
indicate that the magnetic excitations change relatively little,
consistent with maintaining fluctuating stripe correlations
\cite{xu05}.   New optical conductivity \cite{dord05} and photoemission
measurements \cite{vall05} on the same material indicate that stripes are
compatible with the so-called nodal-metal state, but also suggest a
charge-density-wave gap within the stripes that is presumably responsible
for the suppression of superconductivity \cite{kive98}.  This leads to a
new view of stripes and metallic transport.  To appreciate some of these
results, I will first discuss some characteristics of the ``pseudogap''
phase in the next section.

\section{Characteristics of the pseudogap phase}

It is common to describe the normal state of underdoped cuprate
superconductors as a pseudogap phase, at least for temperatures below
some crosssover $T^*$ \cite{timu99}.  There are two types of features
associated with the pseudogap concept.  One is the general depression of
the density of states near the Fermi level, $E_{\rm F}$, in underdoped
cuprates compared to what one might expect for weakly interacting
electrons.  A second is the temperature-dependent depression of signal
seen first in optical conductivity measured with the polarization along
the $c$-axis (perpendicular to the CuO$_2$ planes) \cite{home93} and soon
after in angle-resolved photoemission (ARPES) measurements in the
vicinity of $(\pi,0)$ (the ``antinodal'' region) \cite{loes96,norm98}. 
That the depressed density of states, on the one hand, and the
temperature-dependent effects, on the other, are distinct phenomena
becomes very clear when one considers studies of the optical conductivity
parallel to the CuO$_2$ planes,
$\sigma_{ab}$.

Undoped, insulating cuprates have a charge transfer gap of 1.5 to 2 meV. 
As shown by Uchida {\it et al.} \cite{uchi91} for the case of \lsco, hole
doping introduces finite conductivity into the gap, and this
conductivity grows with doping up to $x\sim0.2$.  (The growth in
integrated conductivity is opposite to what one would expect if the
conduction electrons were noninteracting.)  This effect is visible 
at room temperature; on cooling,
$\sigma_{ab}$ shows no loss of density of states in the normal state
\cite{sant04}.  (Below the superconducting transition temperature, $T_c$,
the contribution from electrons that participate in the superfluid move to
zero frequency, leaving a gap-like feature in $\sigma_{ab}$.) 

In contrast to the 1-eV energy scale for the ``depressed'' density of
states, the temperature-dependent pseudogaps seen in ARPES and $\sigma_c$
have an energy scale on the order of 50 meV, comparable to the 
the superconducting gap maximum.  Clearly, the temperature-dependent
effect occurs on an energy scale that is much smaller than that
associated with the general depression of the density of states.

While no temperature-dependent pseudogaps are seen in $\sigma_{ab}$,
there is nevertheless an interesting variation in $\sigma_{ab}(\omega)$
as $T$ is reduced towards $T_c$.  At high temperatures, the conductivity
of underdoped cuprates is rather flat as a function of frequency
\cite{take03,lee05}.  On cooling, the conductivity below $\sim0.1$~eV
narrows into a Drude peak, while that at higher energies changes little. 
At the lower temperatures it becomes possible to decompose the spectrum
into two components: a narrow Drude peak at low frequency and a broad
Lorentzian oscillator (the ``mid-IR'' peak) with a doping-dependent
mid-point in the range 0.2--0.5~eV \cite{lee05,padi05,rome92}.  In the
superconducting state, the superfluid density comes almost entirely from
the Drude peak \cite{lee05}.  It should be noted that the integrated
conductivity in the Drude peak is only a quarter of that in the mid-IR
\cite{lee05,padi05}.

The Drude peak has to come from states close to $E_{\rm F}$.  In the
underdoped cuprates, these states are along the Fermi arc
\cite{norm98,yosh03}, centered on the nodal point [the point at which the
$d$-symmetry superconducting gap goes to zero, close to
$(\frac{\pi}2,\frac{\pi}2)$].  The regime, below $T^*$, where the Drude
peak can be resolved has been called a nodal-metal state \cite{lee05}.  I
believe that this is a more useful description than the more ambiguous
``pseudogap'' phase.

It was originally noted by Ido {\it et al.} \cite{ido91} that the energy,
magnitude, and doping dependence of the mid-IR peak in \lsco\ is similar
to that in the nickelates, \lsno.  It was only later that the existence of
diagonal stripe order was identified in the nickelates \cite{tran94a}. 
An interpretation of the mid-IR peak in the nickelates in terms of the
electronic structure of stripes has recently been proposed
\cite{home03}.  The main difference between the cuprates and nickelates
is the absence of a Drude component in the latter.  It should be noted,
however, that a Drude peak has been observed in the cuprate
La$_{1.275}$Nd$_{0.6}$Sr$_{0.125}$CuO$_4$ \cite{dumm02}, which should
exhibit stripe order as in similar compositions with slightly less Nd
\cite{ichi00}.  Lorenzana and Seibold \cite{lore03} have obtained both
Drude and mid-IR contributions in a calculation for cuprate stripes.

\section{Universal magnetic excitation spectrum}

As mentioned before, \lbcoate\ exhibits static charge and spin stripe
order for $T<50$~K.  The magnetic excitation has been measured at
$T=12$~K using the MAPS spectrometer at the ISIS Facility \cite{tran04}. 
The dispersion along a line through the positions of a pair of magnetic
superlattice peaks is shown in Fig.~1.  At low energies, the excitations
disperse upwards out of the superlattice peaks as expected for spin waves
\cite{fuji04}.  With increasing energy, we would expect to begin to
resolve cones of spin waves, as observed in stripe-ordered \lsno\
\cite{woo05}; in contrast, we can only identify inwardly dispersing
excitations that merge at about 50 meV.  Above that energy the
excitations disperse outward again, forming an hour-glass shape. 

\begin{figure}
\begin{center}
\includegraphics[height=10cm]{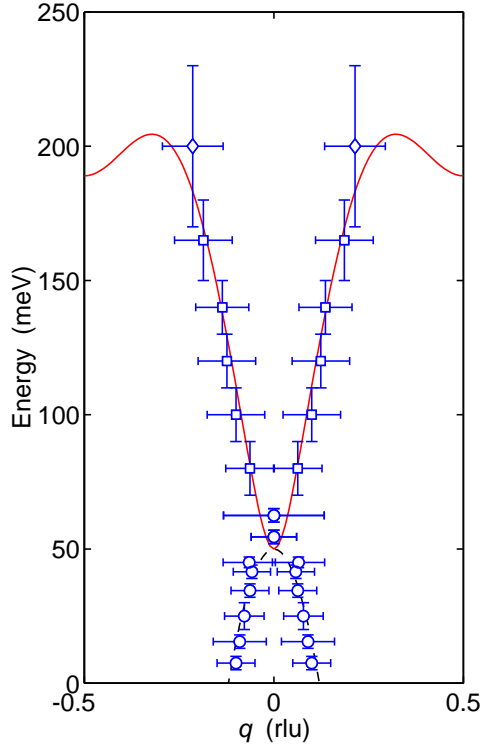}
\end{center}
\caption{ \label{fig:disp} 
Symbols: experimentally measured dispersion\protect\cite{tran04} of
magnetic excitations along ${\bf Q}=(0.5+q,0.5,l)$ in stripe-ordered
La$_{1.875}$Ba$_{0.125}$CuO$_4$.  Horizontal bars indicate the fitted
half-width of the scattering.  The solid line is the
calculated\protect\cite{barn94} dispersion along a two-leg ladder with a
superexchange energy of $J=100$~meV.} 
\end{figure} 

While the observed dispersion differs from the simplest spin-wave
predictions, it turns out to be quite similar to what is observed in
superconducting \ybco\ \cite{arai99,bour00,hayd04,rezn04,stoc04,pail04}
and \lsco\ \cite{chri04}.  The main difference among these systems is
that there is a spin gap in the superconducting state \cite{tran05a}, with
a pile up of weight above the spin gap \cite{chri04,tran04b}.  The
magnitude of the spin gap is roughly proportional to $T_c$ \cite{tran05b}.

Guangyong Xu has recently extended the magnetic excitation measurements
for\linebreak \lbcoate\ to temperatures above the stripe-ordering
transition
\cite{xu05}.  At 65 K, just above the transition, the main differences
are at energies below 10 meV \cite{fuji04,xu05}, and are of the type
expected in a magnetic system that goes through a disordering
transition.  At higher energies, the main effect seems to be a
broadening of the spectrum in terms of momentum widths.  The general
features of the spectrum are still recognizable at 300 K.  The smooth
evolution of the magnetic spectrum suggests that the high-temperature
state is a dynamically-disordered version of the stripe-ordered phase; in
other words, we have evidence for a fluctuating stripe phase.  The
similarity of the spectrum with that of \ybco\ and \lsco\ then suggests
that quantum-disordered stripes may be common to the superconducting
cuprates. 

\section{Electronic nature of the stripe-ordered state}

A remaining question is: are the electronic properties of a
static or fluctuating stripe state compatible with the nodal-metal phase
discussed earlier?  As a first test, Dordevic and Homes \cite{dord05}
have measured the in-plane optical conductivity on a cleaved crystal of
\lbcoate.  The results are generally quite similar to those in the normal
state of \lsco\ with $x=0.125$ \cite{dumm02}.  In particular, the
conductivity below 50 cm$^{-1}$ ($\sim6$~meV) grows with cooling, even in
the stripe-ordered state.  Those low-energy excitations must be associated
with the nodal states along the Fermi arc \cite{yosh03}.  A new behavior
is the gap-like loss of weight from the range 150--300~cm$^{-1}$ on
cooling into the stripe-ordered phase.  This loss of weight is likely to
occur in the antinodal region of reciprocal space.

Tonica Valla, working with Alexei Federov, has very recently succeeded in
measuring angle-resolved photoemission from \lbcoate\ \cite{vall05}. 
Although the measurements were obtained with an energy resolution of
30--35 meV, they provide clear evidence for a Fermi arc and nodal
quasiparticles.  This indicates that stripes are compatible with the
nodal metal; in fact, the main conduction path is at 45$^\circ$ to the
stripes.  

The higher-energy gap-like behavior seen in the optical conductivity
suggests that there might be a charge-density wave (CDW) along the stripes
in the ordered phase.  The competition between CDW and superconducting
correlations within the stripes has been discussed by Kivelson, Fradkin,
and Emery \cite{kive98}, and the nesting of Fermi-surface segments
associated with stripes was noted by Zhou {\it et al.} \cite{zhou99},
with $2k_{\rm F} = a^\ast/4$.  The CDW would require electron-phonon
coupling, and hence one might expect to see a phonon anomaly at 
${\bf q}=(2k_{\rm F},0,0)$ (for stripes running along the $x$
direction).  Intriguingly, Reznik {\it et al.} \cite{rezn05} have
observed a 10-15 meV energy broadening of the longitudinal-optical
bond-stretching branch at just this wave vector.  (Because of the nature
of the crystal structure, the measurement averages over directions
parallel and perpendicular to the stripes, so that it is not possible to
tell directly in which orientation the anomaly occurs.)  Further
measurements are in progress.

\section{New view of stripes and metallic transport}

The new experimental results are changing the picture of metallic
transport in a striped phase.  Previously, the simplest assumption was
that charge would move more easily along a stripe than perpendicular to
it.  The optical conductivity \cite{dord05} and photoemission
\cite{vall05} results now indicate that the best metallic conduction is
at 45$^\circ$ to the stripes.  It appears, then, that stripe correlations
are compatible with the nodal-metal state.  It remains a challenge for
theory to explain these results.


\begin{acknowledgements}

I would like to thank my many experimental collaborators, including
S. Dordevic M. Fujita, G. D. Gu, C. C. Homes, M. H\"ucker, L.
Pintschovius, D. Reznik, T. Valla, G. Xu, and K. Yamada. Research at
Brookhaven is supported by the U.S. Department of Energy's Office of
Science under Contract No.\ DE-AC02-98CH10886.

\end{acknowledgements}



\end{document}